\documentclass[preprint2,tighten]{aastex6}
\usepackage{amsmath,amstext}
\usepackage{amssymb}
\usepackage{gensymb}

\usepackage[T1]{fontenc}
\usepackage{apjfonts} 
\usepackage[figure,figure*]{hypcap}

\shorttitle{Thermal Disk Winds in X-Ray Binaries}
\shortauthors{Higginbottom et al.}

\begin{document}

\title{Thermal Disk Winds in X-ray Binaries: \\ Realistic Heating and Cooling Rates Give Rise to Slow, but Massive Outflows}

\author{N.~Higginbottom\altaffilmark{1}}
\author{D.~Proga\altaffilmark{2}}
\author{C.~Knigge\altaffilmark{1}}
\and
\author{K.~S.~Long\altaffilmark{3,4}}

\altaffiltext{1}{School of Physics and Astronomy, University of Southampton, Highfield, Southampton, SO17 1BJ, UK}
\altaffiltext{2}{Department of Physics \& Astronomy, University of Nevada, Las Vegas, 4505 S. Maryland Pkwy, Las Vegas, NV 89154-4002, USA}
\altaffiltext{3}{Space Telescope Science Institute, 3700 San Martin Drive, Baltimore, MD, 21218, USA}
\altaffiltext{4}{Eureka Scientific Inc., 2542 Delmar Avenue, Suite 100, Oakland, CA, 94602-3017, USA}

\begin{abstract}
A number of X-ray binaries exhibit clear evidence for the presence of disk winds in the high/soft state. A promising driving
mechanism for these outflows is mass loss driven by the thermal expansion of X-ray heated material in the outer disk 
atmosphere. Higginbottom \& Proga recently demonstrated that the properties of thermally-driven winds depend critically 
on the shape of the thermal equilibrium curve, since this determines the thermal stability of the irradiated material. For a given 
spectral energy distribution, the thermal equilibrium curve depends on exact balance between the various heating and 
cooling mechanisms at work. Most previous work on thermally-driven disk winds relied an analytical approximation to these rates.
Here, we use the photoionization code \textsc{cloudy} to generate realistic heating and cooling rates which we then
use in a 2.5D hydrodynamic model computed in ZEUS to simulate thermal winds in a typical black-hole X-ray binary. 
We find that these heating and cooling rates produce a significantly more complex thermal equilibrium curve, with dramatically 
different stability properties. The resulting flow, calculated in the optically thin limit, is qualitatively different from flows 
calculated using approximate analytical rates. Specifically, our thermal disk wind is much denser and slower, with a mass-loss rate
that is a factor of two higher and characteristic velocities that are a factor of three lower. The low velocity of the flow -- $v_{max} \simeq 200$~km~s$^{-1}$ -- may be difficult to reconcile with observations. However, the high mass-loss rate -- 15$\times$  the accretion rate -- is promising, since it has the potential to destabilize the disk. Thermally-driven disk winds may therefore provide a mechanism for state changes.
\end{abstract}

\keywords{accretion, accretion disks - hydrodynamics - methods: numerical - X-rays: binaries}

\section{Introduction}
Disk winds are seen in a range of accreting objects, from cataclysmic variables (CVs) to active galactic nuclei (AGN). 
The existence of these outflows is inferred from the presence of blue-shifted absorption features, 
and detailed analysis of these features can provide information regarding the ionization state, geometry 
and kinematics of the flow.

Amongst the many classes of objects exhibiting such blue shifted absorption lines are X-ray binaries (XRBs). 
These are thought to be systems in which a neutron star (NS) or black hole (BH) is accreting 
 material from a companion star. XRBs are further subdivided into two classes. In high-mass XRBs 
(HMXBs), the compact object has a massive companion and often accretes material from the wind of this early-type star.
By contrast, low-mass XRBs (LMXBs) contain late-type secondaries, and accretion is always driven 
by Roche-lobe overflow from these donors. Our focus here is on LMXBs.

LMXBs are highly variable, exhibiting dramatic changes in their spectral energy distribution (SED). Sometimes they
exhibit a blackbody disk-like spectrum (referred to as the `high/soft' state), and sometimes the SED is dominated by a 
hard power-law spectrum (the `low/hard' state). This second state is often accompanied by a radio jet which disappears as
the source transitions into the high/soft state \citep{2004MNRAS.355.1105F}. The X-ray signatures of
highly ionized gas have been
observed in more than a dozen LMXBs \citep[][and refs therein]{2016AN....337..368D}, with 8 showing evidence of
an outflow.
Outflows are always seen in systems in the high/soft state that are viewed close to edge-on\citep{2012MNRAS.422L..11P}. 
The most likely candidate for this outflowing
gas is an equatorial wind arising from the accretion disk, although the mechanism for launching and accelerating such XRB 
winds is still a subject of active research. 

One promising possibility is that the wind is thermally driven, with the upper surface of the accretion disk puffing up as
it is heated by X-rays from the inner disk and corona. Expansion of this gas as it is further heated then accelerates the
material away from the central object as a wind. Such winds are expected to arise at fairly large distances from the central object,
where the local isothermal sound speed at the Compton temperature exceeds the local escape velocity \citep{1983ApJ...271...70B,
1996ApJ...461..767W}. Alternative wind-driving mechanisms include acceleration via magneto-centrifugal forces \citep[e.g.][]{1982MNRAS.199..883B,1992ApJ...385..460E,2006Natur.441..953M,2016A&A...589A.119C} and radiation pressure acting on electrons or spectral lines \citep[e.g.][]{1980AJ.....85..329I,1993ApJ...409..372S,1985ApJ...294...96S,
2002ApJ...565..455P}.

In order to test these models, we must rely upon measurements of the kinematics of the wind from spectral features. The observationally inferred wind velocities are typcially
between about 400 $\rm{km~s^{-1}}$ and 3000 $\rm{km~s^{-1}}$ across both BH and NS systems
\citep{2016AN....337..368D}, although one system does show a marginal detection of higher velocity gas
\citep{2012ApJ...746L..20K}. Photoionization modelling can also provide estimates of the column density of the 
absorbing gas, along with its ionization parameter. The inferred columns lie between 
about $5\times10^{20}$ and $10^{24}~\rm{cm^{-2}}$ \citep{2016AN....337..368D}, with the plasma in a 
highly ionized state ($\log{\xi}\geq3$). The ionization parameter $\xi$ is defined here as $\xi=L_X/nr^2$ where
$L_X$ is the X-ray luminosity, $r$ is the distance between the gas and the X-ray source, and $n$ is the gas
density.

Focusing on thermally driven disk winds, \citet[hereafter HP15]{2015ApJ...807..107H} took a 
hydrodynamical model of such an outflow first developed by \citet[hereafter L10]{2010ApJ...719..515L} and 
investigated its sensitivity to changes in the assumed heating and cooling rates. 
They found that a reduction in line cooling (which could perhaps reflect optical depth effects) and an
increase in photoionization heating (which could perhaps mimic SED effects) could significantly increase
the mass-loss rate via the thermal wind, from about 4 times the accretion rate to over 40 times. 
Such a high mass-loss rate would be expected to destabilise the disk and could perhaps supply a mechanism for
state change \citep{1986ApJ...306...90S}. In addition, the velocities of the outflowing gas approached
those inferred from X-ray observations.

These results demonstrated that heating and cooling rates are critical in the launching and acceleration 
of thermally driven disk winds. However, the rates used in L10 were analytical approximations to more complex 
underlying rates, and 
the HP15 simply added multiplicative factors to these approximations. Given the sensitivity of the outflow 
properties to heating and cooling found in HP15, we decided to carry out a new set of simulations 
with more detailed and realistic heating and cooling rates.  We continue to use the 
hydrodynamics code \textsc{zeus}  \citep{1992ApJS...80..753S}, extended by \citep{2000ApJ...543..686P}, but now 
we use the radiative transfer and ionization code \textsc{cloudy} \citep{2013RMxAA..49..137F}
to compute heating and cooling rates for the model. This work is part of an ongoing push towards full
radiation hydrodynamic simulations of thermally-driven disk winds that will be able to take full account
of frequency dependant optical depth effects.

\section{Method}
\label{section:method}

Radiative heating enters the equations of hydrodynamics via the energy conservation
equation
\begin{equation}
\rho\frac{D}{Dt}\left(\frac{e}{\rho}\right)=-P \nabla \cdot \mathbf{v}+\rho\mathcal{L},
\label{equation:hydro_energy}
\end{equation}
where $\mathcal{L}$ is the net heating rate per unit mass e is the internal energy density, 
and v is the velocity.  The first term on the right hand side of the
equation represents adiabatic heating/cooling, whilst the second term represents the radiative heating/cooling. Thermal conduction (TC) can also be
an important process in setting the thermal
balance
in X-ray coronae  \citep[e.g.][]{2001ApJ...558..448J}. Therefore, we
computed the TC
contribution to the gas heating/cooling using our solution and compared
this contribution with
all terms in Eq. \ref{equation:hydro_energy}.
We found that in the thermal wind part of our solution, the conduction is
negligible compared to adiabatic cooling and radiative heating. As
expected the conduction rate reaches a maximum where the temperature
increases most rapidly but even there the conduction rate is
two order of magnitude lower than the adiabatic rate. This comparison
indicates that TC has a small effect on the X-ray winds.

In the optically thin limit, for
a given illuminating SED, $\mathcal{L}$ is a function of plasma temperature T, and 
ionization parameter $\xi$, defined here as
\begin{equation}
 \xi=\frac{L_x}{n_Hr^2},
 \label{equation:xi}
 \end{equation}
 where ${L_X}$ is the X-ray luminosity (integrated from 13.6eV to infinity), 
 $n_H$ is the Hydrogen number density of the gas, and r is the distance
 between the source of ionizing flux and the gas. Note that this definition is slightly
 different to that used in HP15. This rescaling is motivated by the aims of our current 
 project, 
 in which we aim to use heating and cooling rates calculated directly from
  photoionization codes rather than from the parametrised equations calculated by
   \citep[hereafter B94]{1994ApJ...435..756B} which were used in HP15. Since the
   ionization parameter is defined in both XSTAR and \textsc{cloudy} as a ratio
   between the ionizing radiation (either energy or photon flux) and the Hydrogen density,  
   choosing this definition of $\xi$ simplifies comparisons between the codes.
 
 The heating and cooling rates parametrised by 
 the `Blondin equations'  from B94 include expressions for net Compton heating ($G_{Compton}$) and 
 bremsstrahlung cooling ($L_b$)
 both derived from first principles, and X-ray heating ($G_X$) and line
 cooling ($L_l$), which are fits to rates calculated from XSTAR simulations \citep{1990ApJ...356..591B}. 
 Line cooling represents
 the energy lost from the plasma to the photon field due to bound-bound and free-bound 
 interactions, whilst X-ray heating represents heating caused by bound-free (photoionization)
 interactions.  These equations enter Equation \ref{equation:hydro_energy}
 via
 \begin{equation}
\rho\mathcal{L}=n_Hn_e(G_{Compton}+G_X-L_l-L_b).
\label{equation:heatcool}
\end{equation}

Whilst the Blondin equations provide reasonable estimates of the heating and cooling rates, particularly for 
Compton and bremsstrahlung effects, the line cooling and X-ray heating rates are fits to 
more complex underlying data. Using these equations is computationally efficient.
However, here we are interested in seeing how the details of the heating and cooling rates affect the
solutions. In addition, the simulations used to generate the fits are some 25 years old, and it is worth
updating the simulations with improved and more complete atomic data and codes.
To achieve this, we compute heating and cooling rates in \textsc{cloudy} from a model
consisting of a thin shell of plasma 
surrounding a point-like
central radiation source. We vary the ionization parameter and temperature in the shell 
in order to produce a grid of heating and cooling rates which are 
supplied to ZEUS via look-up tables. These are used to obtain 
$\mathcal{L}$ at each hydrodynamic time step via bilinear interpolation in $\rm{\log(\xi,T)}$ space. 

In order to compare with the simulations presented by HP15, we use the same SED
that was used to compute the Blondin heating and cooling rates.\footnote{It should be noted that this method permits any SED to be used in 
calculating the heating and cooling rates, and so future simulations using observed SEDs are possible.}
B94 used XSTAR to compute these rates, and the adopted SED was a Bremsstrahlung
spectrum of the form 
\begin{equation}
f_{\nu}\sim \exp(-h\nu/k_BT_X).
\label{equation:xstar_brem}
\end{equation}
The net heating and cooling rates
include all heating and cooling mechanisms included in version 13 of \textsc{cloudy}, with 
the exception of induced Compton heating, which we remove.This mechanism would 
not be expected to be important in a real system, and only appears in our thin shell model
because the \textsc{cloudy} Bremsstrahlung SED does not have a lower frequency cut-off. 
This means that the radiation density
exceeds that of a blackbody at low temperatures. As in HP15, we make the assumption
that the wind is optically thin when computing $\xi$ in each cell.

The simulations are carried out on a spherical polar grid, running from $\rm{0\degree}$ to $\rm{90\degree}$ in 
polar angle 
($\rm{90\degree}$ 
representing the mid-plane) and from $R_{in}$ to $R_{out}$ in radius. In the radial dimension, the spacing between grid
points increases with radius, such that $dr_{k+1}/dr_{k}=1.02$ thereby giving greater resolution in the inner parts of the
flow. Conversely, going from the polar direction towards the disk, the angular grid spacing decreases, such that
$d\theta_{k+1}/d\theta_{k}=0.95$. This gives more resolution near the disk, where we expect any flow will be
accelerating, and therefore changing density, velocity and temperature quickly. A reflecting boundary condition (BC)
is used along the midplane, outflow BCs at the inner and outer radial edges and a rotational BC along the axis of 
rotational symmetry. A density floor of $10^{-22}\rm{g~cm^{-3}}$ is applied throughout the model.

The initial state of the simulation is a hydrostatically supported disk, with the density along the mid-plane set by 
 \begin{equation}
\rho(r)=\rho_0\left(\frac{r}{R_{IC}}\right)^{-2},
\label{equation:midplane_rho}
\end{equation}
where $R_{IC}$ is the Compton radius, i.e. the radius at which the local isothermal sound speed (at the
Compton temperature, $\rm{T_{IC}}$) is equal to the escape velocity,
\begin{equation}
R_{IC} = \frac{GM_{BH}\mu m_H}{k_BT_{IC}}.
\end{equation}
Here, $M_{BH}$ is the mass of the central object and $\mu$ is the mean molecular mass (set to 0.6 in these simulations).
$\rm{R_{IC}}$ is the radius beyond which one would expect a thermal wind
to arise. For a black hole mass of 7$M_{\odot}$, and for the Compton temperature $T_{IC}=1.4\times10^7\rm{K}$,
$\rm{R_{IC}}=\rm{4.8\times10^{11}~\rm{cm}}$. 

\begin{figure}
\includegraphics{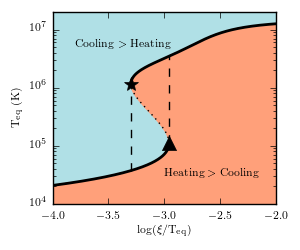}
\caption{An schematic thermal equilibrium curve showing parts of the phase space where heating exceeds 
cooling and vice versa.
The location of $\rm{\xi_{cool,max}}$, the maximum ionization parameter on the cool stable branch is indicated by a triangle, and 
$\rm{\xi_{hot,min}}$, the lowest ionization parameter on the hot stable branch is indicated by a star. The dotted part of the thermal equilibrium curve is unstable.}
\label{figure:simple_stab}
\end{figure}

The choice of the value of $\rho_0$ sets the ionization parameter at the midplane. In order 
to explain the significance of this choice, we briefly discuss the schematic stability curve shown in Figure \ref{figure:simple_stab}.
The familiar `s' shaped curve, formed by the two thick lines connected by the dotted line, represents the points on the 
phase diagram where heating is equal to cooling (the quantity $\rm{\xi/T}$ is inversely proportional to pressure). For gas in 
thermal equilibrium on either of the two thick lines, the equilibrium is {\em stable}. For example, if it is heated 
up in some way, it enters a part of the phase diagram where the cooling rate exceeds the heating rate, so
it will cool back to the equilibrium state. Similarly, if gas is cooled away from from such locations on the thermal 
equilibrium curve, it finds itself in a regime where heating exceeds cooling, restoring the equilibrium. The situation is different for gas in thermal equilibrium on the dotted line. If such a parcel of gas is heated up, it moves into a region where heating exceeds cooling, so it will continue to heat up until it reaches the upper portion of thermal equilibrium curve -- the hot stable branch. Likewise, if gas is cooled it will enter the part of the phase diagram where the cooling rate exceeds the heating rate. It will continue to cool until it reaches the cool stable branch. Thus the dotted part of the S-curve is thermally unstable. 

In our simulations, we are interested in modelling outflowing gas arising from the surface of an accretion disk. 
Points lying on the cool stable branch are in thermal equilibrium at a relatively low temperature, and
do not contribute to any outflow. They can be considered to be a part of a stable accretion disk atmosphere and
we therefore wish to exclude such cells from our simulation.
Maximum heat input to the gas is achieved for gas with an ionization parameter of $\rm{\xi_{cool,max}}$, indicated by a triangle on
Figure \ref{figure:simple_stab}. Assuming that the gas is free to expand, this increase in internal energy will result in 
acceleration of the gas. Gas lying at this point which is heated up slightly has the maximum change in temperature 
and hence change in internal energy before reaching the upper stable branch. 
The temperature of the gas at $\rm{\xi_{cool,max}}$, $\rm{T_{cool,max}}$ is therefore also important. The cooler the
gas is at this point, the greater the potential heat input is, and therefore the greater the acceleration.
In the absence of strong magnetic fields, this change takes place at constant pressure
and the gas parcels will follow  a vertical track on this diagram. 
We therefore set $\rm{\rho_0}$ so that gas at the midplane has this value of $\xi$ and thus can be considered to be the 
uppermost thermally stable surface of
the accretion disk atmosphere. 
The initial temperature for cells along the midplane is set according to the standard thin disk approximation 
\citep{1973A&A....24..337S} with other cells set to the same effective temperature as the midplane cell vertically below
them.

We have carried out two new simulations. Firstly, we use the Blondin equations, with no prefactors applied to compute
the heating and cooling rates - our baseline test. This is broadly comparable to Model A in HP15, and we refer to it
as the `Blondin run'.
Secondly, we have carried out a simulation where the heating and cooling rates are
set using the \textsc{cloudy} lookup table described above. We refer to this run as the `Cloudy run'.

 \begin{figure}
\includegraphics{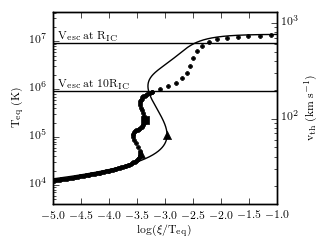}
\caption{Stability curves computed from a thin shell \textsc{cloudy} model (dashed line) and that for the parametrised heating
and cooling rates used in HP15 (solid line). The second y-axis shows the RMS thermal velocity for gas at
these temperatures, and the two horizontal lines show the escape velocity at $\rm{1R_{IC}}$ and $\rm{10R_{IC}}$.}
\label{figure:stab_comp}
\end{figure}

Figure \ref{figure:stab_comp} shows the stability curves for the Blondin equations, and the \textsc{cloudy} simulations.
The location of $\rm{\xi_{cool,max}}$ is indicated on each one by a triangle. 
It is clear that the shape of the \textsc{cloudy} curve is more complex\footnote{See \cite {2016arXiv161004292D} 
for the effect on the stability curve of a range of different SEDs.}, showing an intermediate stable zone at a temperature of
about $2\times10^{5}$ K. Instability sets in once more at the location shown by a square. We also show the isothermal
sound speed on the second axis and the escape velocity at 1 and 10$\rm{R_{IC}}$. We see that gas on the upper stable 
branch has a thermal velocity in excess of the escape velocity at 1$\rm{R_{IC}}$ as expected. 

Another important parameter is the temperature at which gas parcels heating up at constant
pressure from $\rm{\xi_{cool,max}}$ intersect with the stability curve once more,
$\rm{T_{hot,min}}$. The
difference
between this and $\rm{T_{cool,max}}$is a measure of how much energy can be absorbed by the gas as
it heats up in the unstable part of the stability plot. The fact that the value of
$T_{cool,max}$, is lower for the
Cloudy run would tend to suggest that more energy can be absorbed in this case, but the 
presence of the intermediate stable zone and the relatively low value of $\rm{T_{hot,min}}$ means
that this is not the case.

In order to make the comparison a fair one, we set the midplane density in a comparable way for both runs. 
As in HP15, we choose to set the density such that the ionization parameter at the midplane is equal to $\xi_{cool,max}$. \footnote{Note that since both the midplane density and flux drop off as $\rm{r^{-2}}$ the ionization parameter is constant in all midplane cells. This is not to say we suggest that the radial drop off in density is a realistic parametrisation
of the density drop off at any given depth within the disk, rather that we assume that there will be some depth within the disk 
atmosphere at which the ionization parameter will be equal to  $\xi_{cool,max}$. We would expect the denser disk photosphere to exist
below this point.}
For the Blondin run, this is slightly different from the density used in HP15 because of the rescaling of
 $\xi$. This change results in an increase from $\rho_0=1.14\times10^{-12}~\rm{g~cm^{3}}$ to 
$2.75\times10^{-12}~\rm{g~cm^{3}}$. For the Cloudy run $\rho_0=2.03\times10^{-11}\rm{g~cm^{-3}}$.
For all the runs we set $\rm{L_X=3.3\times10^{37}~ergs~s^{-1}}$ which implies an accretion rate of $\rm{\dot{M}_{acc}
=4.4\times10^{17}~g~s^{-1}}$ (assuming an efficiency of 8.3\% for $\rm{M_{BH}=7M_{\odot}}$).

\begin{table}
\begin{tabular}{p{3.0cm}p{0.9cm}p{1.0cm}p{0.9cm}p{1.0cm}}
\hline Physical Parameters & Blondin & Extended Blondin & Cloudy & Extended Cloudy \\ 
\hline
\hline
$\rm{M_{BH}~(M_{\odot})}$ & 7 & 7 & 7 & 7\\
$\rm{T_X~(10^6~}$K) & 56 & 56 & 56 & 56  \\
$\rm{L_X~(10^{37}}~\rm{ergs~s^{-1}})$& 3.3 & 3.3  & 3.3& 3.3 \\
$\rm{\log(\xi_{cold,max})}$&  2.10 & 2.10 & 1.226 & 1.226 \\
$\rm{\log(\Xi_{cold,max})}$&  1.33 & 1.33 &  0.877 & 0.877 \\
$\rm{T_{eq}(\xi_{cold,max})~(10^3~K)}$ & 111 & 111 & 43.7 & 43.7 \\
$\rm{\rho_0~(10^{-12}~g~cm^{-3})}$ & 2.75 & 2.75 & 20.3 & 20.3 \\
$\rm{R_{IC}~(10^{11}~cm)}$ & 4.82 & 4.82 & 4.82 & 4.82  \\
\hline
\multicolumn{5}{l}{Grid parameters}\\
\hline
$\rm{R_{min} (10^{10}~}$cm) & 2.4 & 2.4 & 2.4 & 2.4  \\
$\rm{R_{max} (10^{10}}~$cm) & 96& 960& 96& 960\\
$\rm{R_{disk} (10^{10}}~$cm) & 96& 96& 96& 96\\
$\rm{R_{ratio}}$ &1.05&1.05&1.05&1.05\\
$\rm{N_R}$ & 80& 200&  80& 200\\
$\rm{\theta_{min}}$ & 0.0& 0.0& 0.0& 0.0\\ 
$\rm{\theta_{max}}$ & 90.0& 90.0& 90.0& 90.0\\
$\rm{\theta_{ratio}}$ & 0.95& 0.95& 0.95& 0.95\\
$\rm{N_{\theta}}$ & 100& 100& 100& 100\\
\hline
\multicolumn{5}{l}{Derived wind properties}\\
\hline 
\multicolumn{5}{l}{$\rm{V_r~(\rho>1\times10^{12})}$}  \\
$\rm{max,blueshifted~({km~s^{-1}})}$& 79 &  78 &  21 &  19  \\
$\rm{n_H~(60\degree)~(10^{22}}~\rm{cm^{-2})}$ & 2.9 & 3.3 & 3.6 & 4.7  \\
$\rm{n_H~(80\degree)~(10^{22}}~\rm{cm^{-2})}$ & 16 & 17 & 62 & 63  \\
$\rm{\dot{M}_{wind,outer}~(10^{18}~g~s^{-1}})$ & 3.4 &  3.6 &  2.3 &  6.7  \\
$\rm{\dot{M}_{wind,outer}~(\dot{M}_{acc})}$ & 7.8 &  8.2 &  5.2 &  15  \\

\hline
\end{tabular}
\caption{The heating and cooling parameters adopted in the simulations, and
some key parameters of the resulting winds.}
\label{table:wind_param}
\end{table}

\section{Results}
\label{section:results}

\begin{figure*}
\gridline{\leftfig{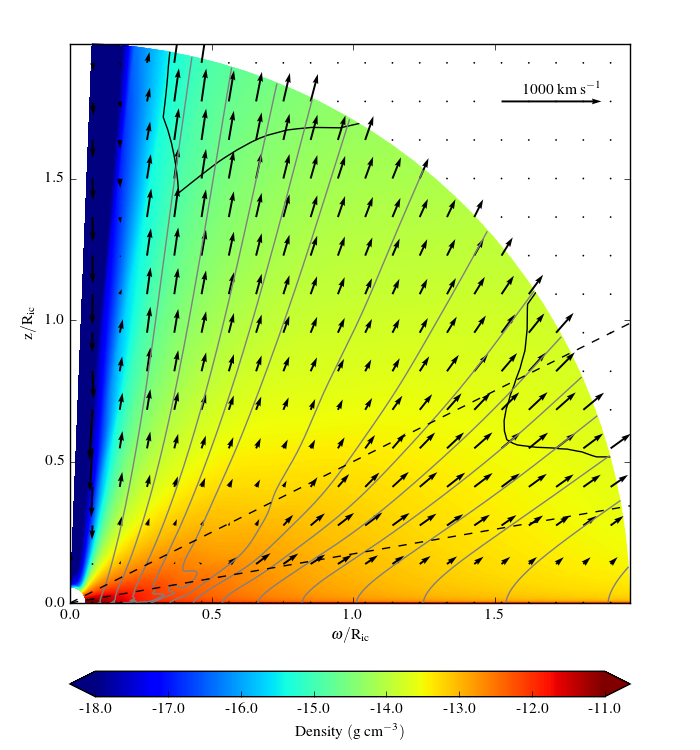}{0.5\textwidth}{}
\rightfig{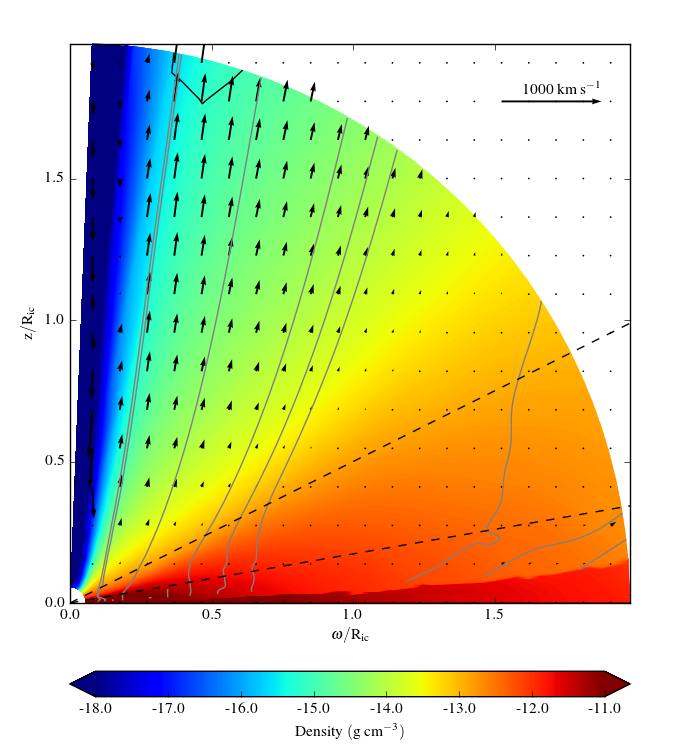}{0.5\textwidth}{}}
\gridline{\leftfig{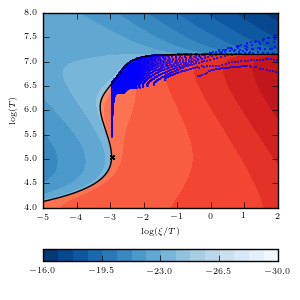}{0.5\textwidth}{}
\rightfig{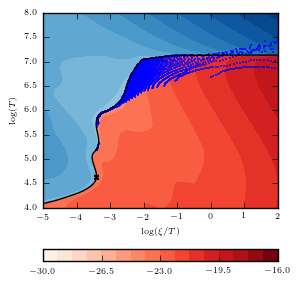}{0.5\textwidth}{}}

\caption{Comparison between two wind geometries. Left hand panels are for Blondin run, right hand panels for
Cloudy run. 
Grey lines show streamlines, and the black line shows the location of the Mach1 surface (seen as two 'v' shaped
features at the outer edge of the Blondin run, and only barely visible at the top of the Cloudy run). The two dotted
lines show the location of the 60\degree and 80\degree sightlines
The two lower panels show the location of all the cells in each model on a phase diagram. The
contours show the net heating/cooling rate and the equilibrium curve is plotted as a thick black line.}
\label{figure:wind_images}
\end{figure*}

The top two panels of Figure
\ref{figure:wind_images} show the density and velocity structure of the Blondin run (left)
compared to Cloudy run (right). It is immediately
clear that the equatorial flow is much slower in the new run, although it is denser. 
We compute a mass outflow rate through the outer simulation boundary by performing the summation
\begin{equation}
\dot{M}=\sum{v_r\rho dA}
\end{equation}
over all of the outer radial cells.
As we would expect,  the Blondin run has a very similar structure to model A in HP15, although the outflow 
 rate is doubled to about $3.4\times10^{18}~\rm{g~s^{-1}}$ (8~$\rm{\dot{M}_{acc}}$). 
 This is because the midplane density in the new simulation is roughly twice that of the older
run for the same value of $\xi$ because of the revised definition of $\xi$.

The Cloudy run has a somewhat variable outflow rate with time, failing to settle down to a steady state even after
$10^7$  seconds of simulation time (which represents at least 32 sound crossing times, calculated by dividing
the radial extent of the model by the slowest isothermal sound speed seen in any cell). 
The mean mass outflow rate over the last $10^6$s is
$2.3\times10^{18}~\rm{g~s^{-1}}$ (5~$\rm{\dot{M}_{acc}}$) with a standard deviation of 
$2.9\times10^{17}~\rm{g~s^{-1}}$. This instability suggests that in this run our domain size may not be large 
enough to capture any stable outflow, and we are instead only simulating the turbulent `corona' which exists
in the original model, but only in the inner $\rm{0.5~R_{IC}}$.

There is a clear density `step' at around 85\degree in the Cloudy run. This is caused by the intermediate stable zone
on the stability curve.  Previous studies have noted the presence of intermediate stable zones. For example one of the stability curves used by \cite{2000ApJ...537..833N} had an intermediate zone which could `interrupt' the heating of plasma from the cool stable branch to the hot stable branch. However an investigation of the effect on gas dynamics was beyond the scope of their study. The standard AGN SED used by \cite{1996ApJ...461..767W} also gave rise to a stability curve with an intermediate  stable zone. In this case,  because the intermediate unstable zone exists at a lower value of $\rm{\xi/T}$ than  $\rm{\xi_{cool,max}/T}$, it is not possible for gas heating up from the cool branch to reach this area of the stability curve and so 
there is no corresponding density step.

The lower two panels of Figure \ref{figure:wind_images} shows where all cells in each 
model lies on a phase diagram. Also plotted are contours of heating and cooling rate and the stability curve
where the heating and cooling rates are equal. The left hand panel shows the familiar behavior of a simulation
with a `normal' S curve with just one unstable zone. The black crosses show the location of cells in the midplane, 
and as expected they all lie at the same point; the last stable point on the lower temperature branch. There is
then a vertical line of cells showing cells heating up at constant pressure just above the midplane. Adiabatic
cooling (and sometimes heating) explains why cells do not always lie on the stability curve. 

In contrast, in the Cloudy run, there is a second cluster of cells around the intermediate stable zone on the 
stability curve (circled in the lower right panel of Figure \ref{figure:wind_images}). These are the dense cells seen 
on the contour plot. Because the cells are on a stable part of the
stability curve, they heat up only slowly since radiative cooling is able to almost balance the heating. An outflow does not 
form. Once the gas heats up enough to reach the upper limit of the second stable zone, they heat up more 
quickly; however the start of the upper stable branch occurs at a much lower temperature than in the original 
simulation (around $6\times10^5$K vs $3\times10^6$K). Since the unstable heating process starts at
a higher temperature, there is significantly less energy deposited into the gas, and so the final outflow
is much slower and unstable.

In order to see if a stable outflow forms at larger radii, we next extend the simulation domain size. However 
this requires care, since the accretion disk in an LMXB has to fit within the primary's Roche lobe. 
An estimate of the maximum disk size in a binary system is provided by the radius of the last non-intersecting particle orbit 
around the accretor. This is given by 
\begin{equation}
\frac{r_d(max)}{a}=\frac{0.6}{1+q}
\end{equation}
where q is the secondary mass divided by the primary, and a is the separation of the binary system \citep{2003cvs..book.....W}. Reliable system parameters
are available for two black-hole binaries that show clear evidence for disk winds in the soft state \citep{2012MNRAS.422L..11P}:
GROJ1655-40 and GRS1915+105. Using the data from \cite{2003A&A...404..301R},
we calculate maximum stable disk radii of$4.7\times10^{11}$ and $3.8\times10^{12}$~cm, respectively. 
Our initial runs with an outer boundary at $\rm{2R_{IC}=9.6\times10^{11}}$cm therefore represent a reasonable
maximum disk radius; when extending the domain, we need to modify the simulation so that the disk does not exceed the size
of the binary.

In order to achieve this, the initial density in the new expanded runs is set along the midplane according to the equation 
\ref{equation:midplane_rho}
up to a radius of $\rm{2R_{IC}}$. Outside this radius, it is multiplied by a factor of $\rm{\exp(20(1-(r/2R_{IC})))}$ to produce
a steep but not instant dropoff to the simulation minimum density of $1\rm{\times10^{-22}~g~cm^{-3}}$

The density at the midplane up to $\rm{r=2R_{IC}}$ is reset each cycle, but the midplane 
density at larger radii is allowed to vary, with the same density floor 
applied as in the rest of the model. The boundary condition varies along the midplane, with the normal reflecting 
boundary condition set out to a radius of $\rm{2R_{IC}}$ but an outflow condition set at larger radii. The fate
of material flowing 'downwards' through the outer disk is not computed, little mass is actually lost 
through this boundary so this somewhat unphysical situation does not significantly affect the results. We compute
two new runs with these parameters, the parameters and some results are summarised in Table 
\ref{table:wind_param}. They are labeled "Extended Blondin" and "Extended Cloudy".

\begin{figure*}
\gridline{\leftfig{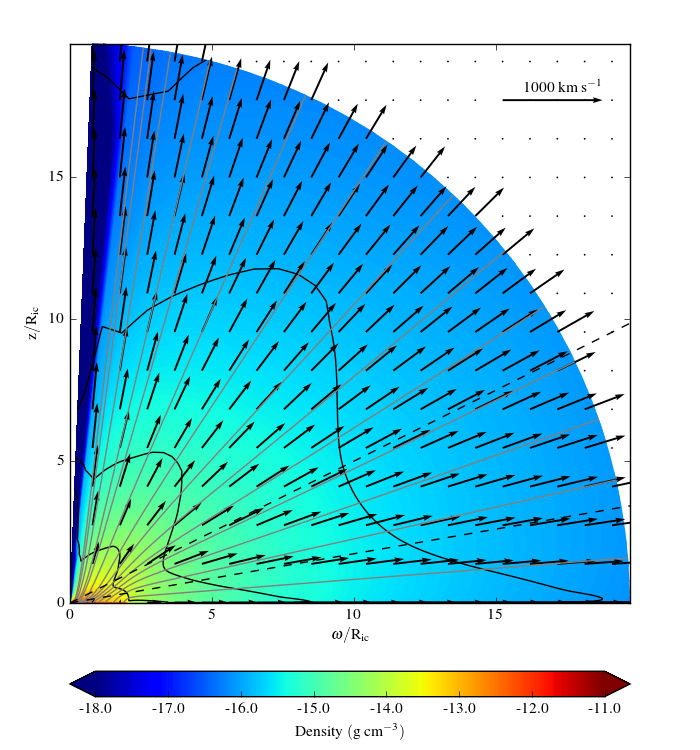}{0.5\textwidth}{}
\rightfig{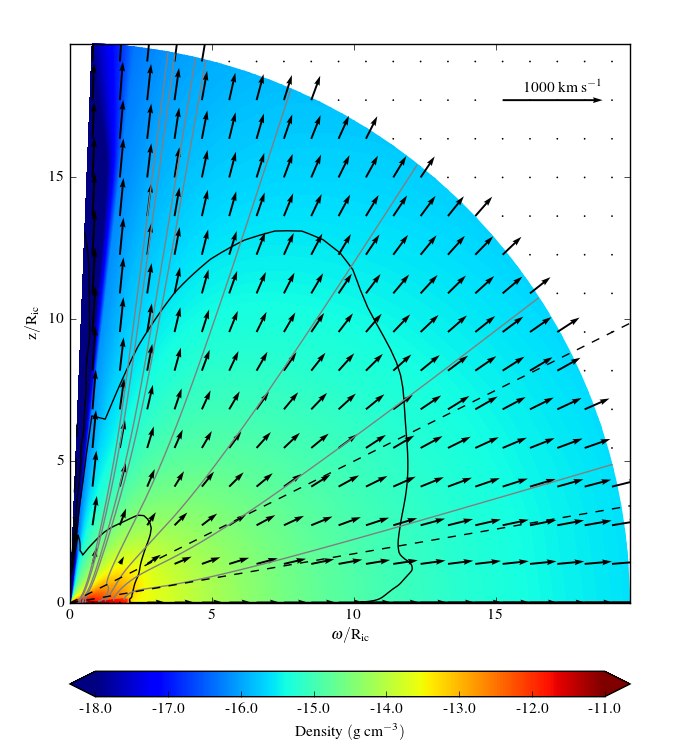}{0.5\textwidth}{}}

\caption{Comparison between two wind geometries with extended simulation domain. 
The left hand panel shows the extended Blondin simulation, the right hand panel shows the extended Cloudy
run. 
Grey lines show streamlines, and the black line shows the location of the Mach surfaces. In each case, the
innermost line  (seen inside about $\rm{2R_{IC}}$) is the Mach 1 surface. The two dotted
lines show the location of the 60\degree and 80\degree sightlines.}
\label{figure:wind_images2}
\end{figure*}

\begin{figure}
\gridline{\leftfig{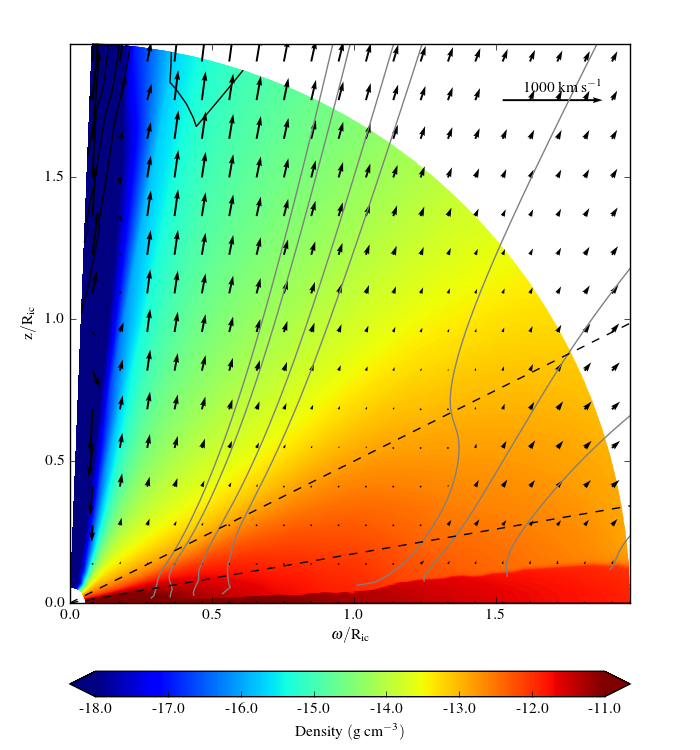}{0.5\textwidth}{}}
\caption{Inner $2\rm{R_{IC}}$ of the extended domain simulation using \textsc{cloudy}
heating and cooling rates}
\label{figure:dw40_inner}
\end{figure}

Figure \ref{figure:wind_images2} shows the density and velocity distribution of the two extended runs. 
The gas continues to accelerate in both cases, but the extended Cloudy run is
still slower than the Blondin version. It is initially surprising to note that the mass loss through
the outer boundary in the extended Cloudy run has significantly increased over the 
smaller domain. The reason is most likely because we do not capture the Mach 1 surface
in the small domain, and so the conditions in the larger run can affect the conditions at the 
base of the wind. This effect is much less in evidence in the Blondin runs, because we do 
include most of the Mach 1 surface in the smaller domain. Figure \ref{figure:dw40_inner}
shows only the inner 2$\rm{R_{IC}}$ of the extended Cloudy run, and it is clear that there are differences
compared to the simulation conducted in the smaller domain. The mass loss seen in the
extended Cloudy run, $15\rm{\dot{M}_{acc}}$, is of the right order of magnitude to cause instabilities
in the disk and perhaps cause state change \citep{1986ApJ...306...90S}.

\section{Discussion}

Whilst the overall mass loss rate is significant, it is important to ask whether this wind would be
observable. 
All XRBs identified as having outflows
exhibit Fe\textsc{xxvi} absorption \citep{2012MNRAS.422L..11P} Some also have Fe\textsc{xxv} lines
and the important density diagnostic 11.770/11.920\AA~ forbidden lines of Fe\textsc{xxii} are seen in at least one 
system \citep{2006Natur.441..953M}. The typical Hydrogen
column density derived from photoionization modelling of lines seen in outflows is between 
$5\times10^{20}$ and $\rm{10^{24}~\rm{ cm^{-2}}}$ \citep{2016AN....337..368D}. The ionization
parameter calculated from the same modelling is in the range $\log(\xi)\sim1.8-6$ 
\citep{2013AcPol..53..659D} with most systems characterised by $\log{\xi} \geq 3$

\begin{figure}
\gridline{\leftfig{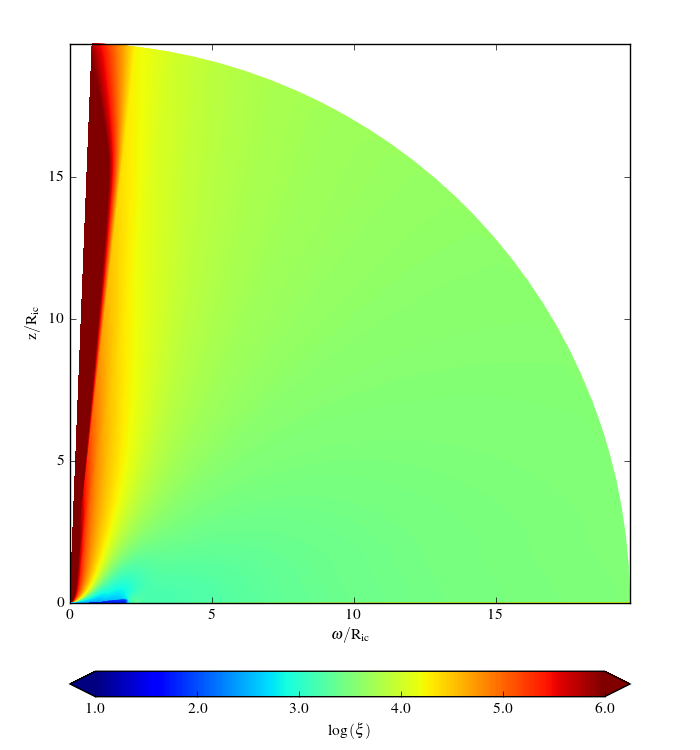}{0.5\textwidth}{}}
\caption{Ionization parameter recorded in extended cloudy model}
\label{figure:IP}
\end{figure}

Figure \ref{figure:IP} shows the value of $\rm{\xi}$ calculated in each cell in the extended Cloudy model 
(still in the optically thin limit) and it confirms that almost the entire model contains gas with an ionization
parameter in the range $\rm{3< log(\xi) <4}$, in the middle of the range of observed ionization parameters.

\begin{figure}
\includegraphics{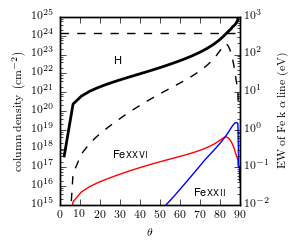}
\caption{Column density of Hydrogen and two Iron species through the extended cloudy model.
 The horizontal dashed 
line shows the value of $\rm{1/\sigma_T}$, and the dashed curve (plotted on the right hand axis) shows the
predicted equivent width of an Iron K$\rm{\alpha}$ absorption feature}
\label{figure:column_density}
\end{figure}

Figure \ref{figure:column_density} shows the column density through the model for Hydrogen and the 
two of the observed Iron species. We see
immediately that the model is somewhat Compton thick for angles greater than about 84\degree. Almost all
sightlines have a Hydrogen column density in the observed range, with more equatorial sightlines matching 
the higher estimates.

Figure \ref{figure:column_density} also shows the column density of these three iron species through the model.
The relative abundances of these ions is computed using the same lookup table method as we used to
calculate the heating and cooling rates. The same thin shell \textsc{cloudy} model is used to produce these
abundance tables.
In the optically thin limit, an estimate of the equivalent width of lines can be obtained from \citep{1994MNRAS.267L..17M}

\begin{equation}
EW=18.1\left(\frac{A_{Fe}}{3.3\times10^{-5}} \right)\left(\frac{N_H}{10^{22}cm^{-2}}\right)\left(\frac{f_{lu}}{0.5}\right)f_{ion}~eV.
\label{equation:EW}
\end{equation}
where $A_{Fe}$ is the iron abundance relative to Hydrogen.
Treating the Iron $K\alpha$ absorption doublet as a single line with oscillator strength $f_{lu}=0.46$ and assuming all
ions are in the lower level we
can estimate the equivalent width of this line as seen from different angles. This is plotted on figure
\ref{figure:column_density}as the dashed curve. Where outflows are detected, they are typically at the level of
10eV and greater \cite{2012MNRAS.422L..11P} so based upon this simple estimate, we would expect this wind
to be detectable for viewing angles of about 50\degree and greater.

Whilst this simple calculation demonstrates that that the simulation produces a gas distribution that would
be detectable, at least in the Iron K$\alpha$ absorption line, it does not demonstrate that it
would produce the blue-shifted features that suggest outflows. Indeed the velocity profiles shown in 
figures \ref{figure:wind_images}-\ref{figure:dw40_inner} show that the typical velocities in the Cloudy models
are 100-200$\rm{kms^{-1}}$ with the dense parts of the wind ($\rm{n_H>1\times10^{12}~cm^{-3}}$) showing outflow 
velocities of less than 100$~\rm{kms^{-1}}$. Typical velocities seen in LMXB outflows are between 
400 and 3000 $\rm{kms^{-1}}$ \citep{2016AN....337..368D}, so the wind produced here is certainly at the 
lower end of observed velocities.
The production of detailed synthetic
spectra is beyond the scope of this work. However, we can compute a simple line profile for the Iron K$\alpha$
 absorption feature that we have just demonstrated is likely to be visible. 

The absorption coefficient for a spectral line, neglecting collisional de-excitation is given by 
\begin{equation}
\alpha(\nu)=\frac{\pi e^2}{m_ec}\phi(\nu)n_lf_{lu},
\label{equation:alpha}
\end{equation}
where $\rm{n_l}$ is the number density of the lower ion and $\rm{\phi{\nu}}$ is the line profile function, which we treat as a 
thermally broadened Gaussian. For each radial cell, we compute the line opacity as a function of frequency, 
Doppler shifted to the radial velocity of that cell. We then compute a total, frequency dependent optical
depth for a given sightline by performing a summation over each cell i, with absorption coefficient $\rm{\alpha_i(\nu)}$
and radial extent 
$\rm{\Delta r_i}$ in that sightline;
\begin{equation}
\tau(\nu)=\sum_{i=inner}^{i=outer}\alpha_i(\nu)\Delta r_i.
\end{equation}
We then produce a synthetic line profile using 
\begin{equation}
F(\nu)=e^{-\tau(\nu)}
\end{equation}

\begin{figure}
\includegraphics{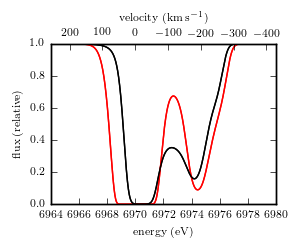}
\caption{Synthetic Iron K$\alpha$ lines at 60\degree (black) and 80\degree(red) for the extended Cloudy
simulation}
\label{figure:line_profile}
\end{figure}

Figure \ref{figure:line_profile} shows the resulting line profiles for two viewing angles. We see that there are two features at
each angle. There are black absorption features at the rest velocity, somewhat narrower at 60\degree than
at 80\degree. This is produced by material at small radii, where the gas is dense but slow moving. The fact that the 
lines are saturated means that strictly speaking Equation \ref{equation:EW} is not universally applicable in the model; We would 
expect the EW to reach a maximum value as the lines saturate. Nonetheless, the statement that the wind should produce observable 
lines is still reasonable although the actual values of EW shown on Figure \ref{figure:column_density} should be taken as upper
limits.
The figures in Table \ref{table:wind_param} for the radial velocities of the high density plasma in the 
simulations underline the fact that where the gas is dense, it is slow moving. There are only two
observed systems where densities of the absorbing gas have been estimated \citep{2008ApJ...672.1091S,
2009ApJ...701..865K}  and in those cases fast moving 
(around $400~\rm{km~s^{-1}}$)
gas with number densities of $10^{13}-10^{15}\rm{~cm^{-3}}$ was observed. Our maximum radial blue shifted velocities
for gas with density above $10^{12}\rm{~cm^{-3}}$ are less than $100~\rm{km~s^{-1}}$ in all cases, slower in the
two Cloudy runs than in the Blondin runs. 
There
is a second feature at about $200~\rm{km~s^{-1}}$ produced by the more tenuous accelerated gas at larger
radii. This underlines the fact that this outflow is significantly slower than most reported in the literature. It is 
interesting to note, however, that the mass loss rate is still significant, confirming the results of \cite{2010ApJ...719..515L}
and HP15.

Our next project is to compute the heating and cooling rates using a full treatment of radiative transfer (RT).
However, it is instructive to calculate whether it is likely that this will have an effect. We can assess this by  checking
whether the plasma at the base of the wind is optically thin
to its own line radiation. This is important, since line cooling is the dominant cooling mechanism at the 
base of the wind.  If the line radiation is unable to escape, then the thermal balance will change significantly.

The optical depth of a spectral line in an expanding spherical atmosphere is given by \citep{1975ApJ...195..157C}

\begin{equation}
\tau_L=\kappa_L\rho v_{th}\left|\frac{dv}{dr}\right|^{-1},
\end{equation}

where  $\rm{v_{th}}$ is the thermal velocity of the gas, and the final term is the velocity gradient.  This equation
is a form of the Sobolev approximation.
$\kappa_L$ is the monochromatic line absorption coefficient, per unit mass, divided by the
line profile function ($\Delta v_D=\nu_0v_th/c$), 

\begin{equation}
\kappa_L=\frac{\pi e^2}{mc}g_1f\left(\frac{n_1}{g_1}-\frac{n_2}{g_2}\right)\frac{1}{\Delta v_D}\frac{1}{\rho}
\end{equation}

Substituting for $\kappa_L$, assuming all ions are in the ground state ($n_2=0$) and simplifying we obtain

\begin{equation}
\tau_L=\frac{\pi e^2}{mc}n_1f\frac{c}{\nu_0}\left|\frac{dv}{dr}\right|^{-1}.
\label{equation:line_tau}
\end{equation}

Line cooling dominates the cooling rate at the midplane of the simulation, and
the main cooling line for $\rm{\xi=\xi_{cool,max}=15}$ and $\rm{T=T_{cool,max}=4.3\times10^{4}~K}$ is  
O~\textsc{vi} $\rm{\lambda 1032/1038~}$\AA,
making up about 30\% of the total cooling rate. Taking this doublet as an example, we can attempt to quantify the effect of
line optical depth on the cooling rate. The oscillator strengths are $\rm{f=0.13,0.07}$ respectively, we treat them as a single line with f=0.2. 
At $\rm{r=R_{IC}}$, $n_1=\rm{n_{O\textsc{vi}}=6.7\times10^8~cm^{-3}}$, the vertical velocity gradient at the midplane is very variable, we take a representative value of $\rm{dv_{z}/dz=2\times10^{-4}~s^{-1}}$ (the radial velocity gradient is zero by construction at the midplane, and the Keplerian velocity shear is an order of magnitude lower than the vertical velocity gradient). These values give an optical depth of about $2\times10^{5}$, highly optically thick.

The 
effect of this is complex. The line cooling will likely be suppressed, however even in a region with such a large optical depth, line radiation can still escape after many scatters. The degree to which line cooling is suppressed depends on the chance of an ion collisionally de-exciting during the scattering process. The effect of this is to reduce the line cooling rate by a factor of $\rm{A_{21}/\tau n_eq_{21}}$ when the optical depth is large \citep[and private communication]{1993ApJ...412..267R}. $\rm{A_{21}}$ is the Einstein radiative recombination rate and $\rm{q_{21}}$ is the collisional de-excitation rate. For
the  O~\textsc{vi} line, this gives a 45 fold suppression factor at $\rm {r=R_{IC}}$. Although this factor decreases with radius  there will certainly be an effect on the wind solution; HP15 experimented with suppressing line cooling by a factor of 5, and this (together with a four-fold increase in X-ray heating) significantly increased the mass outflow rate and velocity of the wind.
Naturally, the any changes to the flow will modify the density, temperature and therefore the suppression factor which will
in turn modify the flow. The forthcoming full RT simulations will capture this behavior.

\section{Conclusions}
\label{section:conclusions}

HP15 demonstrated that the heating and cooling rates are critical in defining
the velocity and density of thermal winds in LMXBs, as calculated in hydrodynamical simulations. Modest 
changes to those rates produced
winds with velocities approaching those seen in observations of LMXBs. The changes 
made to the rates were physically motivated but they were not calculated from
the illuminating SED, plasma composition or any other physical properties of the systems.

In this study, we replaced the parametrized heating and cooling curves used by HP15 with more realistic heating and cooling curves, as  calculated with \textsc{cloudy}. In our hydrodynamical simulations using these rates, the resulting outflow is slower ($\rm{v\lesssim 200~km~s^{-1}}$) than in the simulations carried out by HP15. It is also slower than the wind velocities typically observed in LMXBs, which lie between 400 and 3000 $\rm{kms^{-1}}$ \citep{2016AN....337..368D},  However, the overall mass-loss rate of the outflow 
is significant ($\dot{M}_{wind} \simeq 15~\dot{M}_{acc}$) and may be enough to destabilize the disk and 
perhaps prompt a state change. 

In addition, the wind arises at 
rather large distances from the central object, at least an order of magnitude greater than the 
0.1$\rm{R_{IC}}$ seen in previous studies 
\cite[e.g.][]{1983ApJ...271...70B,1996ApJ...461..767W,2002ApJ...565..455P}. This is because
gas in the new simulations heats up more slowly as a result of intermediate stable temperatures 
existing on the \textsc{cloudy} calculated stability curve, meaning only at larger radii does the gas gain 
sufficient
thermal energy to escape the gravitational potential. This 
is in line with the observation that outflowing material in LMXBs is only seen in long period systems
\citep{2016AN....337..368D}, although \cite{2006Natur.441..953M} found that the outflowing material
in GRO J1655-40 was best modelled by an absorber at about 0.001$R_{IC}$.

\section{Future work}
\label{section:future}
The simulations presented here are all calculated in the optically thin limit. In other words the calculation
of the ionization parameter used to compute the heating and cooling rates in each cell takes no
account of absorption between the central X-ray source and the cell in question. Since heating takes
place in all cells, this is incorrect in detail. In fact, we have shown that the cells closest to
the disk in our simulation are likely to be optically thick to their own line cooling radiation, and so
our next step is to use a full RT
code to compute the heating and cooling rates in each cell, thereby taking account of all absorption 
processes. This is a significantly more complex undertaking and will be described in a following 
paper.

\acknowledgments
Calculations in this work made use of the Iridis4 Supercomputer at the University of Southampton.
DP acknowledges support provided  by NASA under ATP grant NNX14AK44G.  NH and CK  acknowledge support by the Science and Technology Facilities Council grant ST/M001326/1. KSL acknowledges the support of NASA for this work through grant NNG15PP48P to serve as a science adviser to the Astro-H project
We would also like to thank Stuart Sim and James Matthews for many helpful suggestions to
improve this work. Finally we thank the referee
John Raymond for a very useful review.
\software{ZEUS \citep{1992ApJS...80..753S}, CLOUDY \citep{2013RMxAA..49..137F}}
\bibliographystyle{yahapj}
\bibliography{references}

\begin{thebibliography}{}
\providecommand\natexlab[1]{#1}
\providecommand\JournalTitle[1]{#1}

\bibitem[{{Begelman} {et~al.}(1983){Begelman}, {McKee}, \&
  {Shields}}]{1983ApJ...271...70B}
{Begelman}, M.~C., {McKee}, C.~F., \& {Shields}, G.~A. 1983,
  \href{http://dx.doi.org/10.1086/161178}{\JournalTitle{\apj}, 271, 70}

\bibitem[{{Blandford} \& {Payne}(1982)}]{1982MNRAS.199..883B}
{Blandford}, R.~D., \& {Payne}, D.~G. 1982,
  \href{http://dx.doi.org/10.1093/mnras/199.4.883}{\JournalTitle{\mnras}, 199,
  883}

\bibitem[{{Blondin}(1994)}]{1994ApJ...435..756B}
{Blondin}, J.~M. 1994,
  \href{http://dx.doi.org/10.1086/174853}{\JournalTitle{\apj}, 435, 756}

\bibitem[{{Blondin} {et~al.}(1990){Blondin}, {Kallman}, {Fryxell}, \&
  {Taam}}]{1990ApJ...356..591B}
{Blondin}, J.~M., {Kallman}, T.~R., {Fryxell}, B.~A., \& {Taam}, R.~E. 1990,
  \href{http://dx.doi.org/10.1086/168865}{\JournalTitle{\apj}, 356, 591}

\bibitem[{{Castor} {et~al.}(1975){Castor}, {Abbott}, \&
  {Klein}}]{1975ApJ...195..157C}
{Castor}, J.~I., {Abbott}, D.~C., \& {Klein}, R.~I. 1975,
  \href{http://dx.doi.org/10.1086/153315}{\JournalTitle{\apj}, 195, 157}

\bibitem[{{Chakravorty} {et~al.}(2016){Chakravorty}, {Petrucci}, {Ferreira},
  {Henri}, {Belmont}, {Clavel}, {Corbel}, {Rodriguez}, {Coriat}, {Drappeau}, \&
  {Malzac}}]{2016A&A...589A.119C}
{Chakravorty}, S., {Petrucci}, P.-O., {Ferreira}, J., {et~al.} 2016,
  \href{http://dx.doi.org/10.1051/0004-6361/201527163}{\JournalTitle{\aap},
  589, A119}

\bibitem[{{D{\'{\i}}az Trigo} \& {Boirin}(2013)}]{2013AcPol..53..659D}
{D{\'{\i}}az Trigo}, M., \& {Boirin}, L. 2013, \JournalTitle{Acta
  Polytechnica}, 53, 659

\bibitem[{{D{\'{\i}}az Trigo} \& {Boirin}(2016)}]{2016AN....337..368D}
---. 2016,
  \href{http://dx.doi.org/10.1002/asna.201612315}{\JournalTitle{Astronomische
  Nachrichten}, 337, 368}

\bibitem[{{Dyda} {et~al.}(2016){Dyda}, {Dannen}, {Waters}, \&
  {Proga}}]{2016arXiv161004292D}
{Dyda}, S., {Dannen}, R., {Waters}, T., \& {Proga}, D. 2016,
  \JournalTitle{ArXiv e-prints},
  \href{http://arxiv.org/abs/1610.04292}{{\sffamily arXiv:1610.04292
  [astro-ph.HE]}}

\bibitem[{{Emmering} {et~al.}(1992){Emmering}, {Blandford}, \&
  {Shlosman}}]{1992ApJ...385..460E}
{Emmering}, R.~T., {Blandford}, R.~D., \& {Shlosman}, I. 1992,
  \href{http://dx.doi.org/10.1086/170955}{\JournalTitle{\apj}, 385, 460}

\bibitem[{{Fender} {et~al.}(2004){Fender}, {Belloni}, \&
  {Gallo}}]{2004MNRAS.355.1105F}
{Fender}, R.~P., {Belloni}, T.~M., \& {Gallo}, E. 2004,
  \href{http://dx.doi.org/10.1111/j.1365-2966.2004.08384.x}{\JournalTitle{\mnras},
  355, 1105}

\bibitem[{{Ferland} {et~al.}(2013){Ferland}, {Porter}, {van Hoof}, {Williams},
  {Abel}, {Lykins}, {Shaw}, {Henney}, \& {Stancil}}]{2013RMxAA..49..137F}
{Ferland}, G.~J., {Porter}, R.~L., {van Hoof}, P.~A.~M., {et~al.} 2013,
  \JournalTitle{\rmxaa}, 49, 137

\bibitem[{{Higginbottom} \& {Proga}(2015)}]{2015ApJ...807..107H}
{Higginbottom}, N., \& {Proga}, D. 2015,
  \href{http://dx.doi.org/10.1088/0004-637X/807/1/107}{\JournalTitle{\apj},
  807, 107}

\bibitem[{{Icke}(1980)}]{1980AJ.....85..329I}
{Icke}, V. 1980, \href{http://dx.doi.org/10.1086/112678}{\JournalTitle{\aj},
  85, 329}

\bibitem[{{Jimenez-Garate} {et~al.}(2001){Jimenez-Garate}, {Raymond},
  {Liedahl}, \& {Hailey}}]{2001ApJ...558..448J}
{Jimenez-Garate}, M.~A., {Raymond}, J.~C., {Liedahl}, D.~A., \& {Hailey}, C.~J.
  2001, \href{http://dx.doi.org/10.1086/322465}{\JournalTitle{\apj}, 558, 448}

\bibitem[{{Kallman} {et~al.}(2009){Kallman}, {Bautista}, {Goriely}, {Mendoza},
  {Miller}, {Palmeri}, {Quinet}, \& {Raymond}}]{2009ApJ...701..865K}
{Kallman}, T.~R., {Bautista}, M.~A., {Goriely}, S., {et~al.} 2009,
  \href{http://dx.doi.org/10.1088/0004-637X/701/2/865}{\JournalTitle{\apj},
  701, 865}

\bibitem[{{King} {et~al.}(2012){King}, {Miller}, {Raymond}, {Fabian},
  {Reynolds}, {Kallman}, {Maitra}, {Cackett}, \& {Rupen}}]{2012ApJ...746L..20K}
{King}, A.~L., {Miller}, J.~M., {Raymond}, J., {et~al.} 2012,
  \href{http://dx.doi.org/10.1088/2041-8205/746/2/L20}{\JournalTitle{\apjl},
  746, L20}

\bibitem[{{Luketic} {et~al.}(2010){Luketic}, {Proga}, {Kallman}, {Raymond}, \&
  {Miller}}]{2010ApJ...719..515L}
{Luketic}, S., {Proga}, D., {Kallman}, T.~R., {Raymond}, J.~C., \& {Miller},
  J.~M. 2010,
  \href{http://dx.doi.org/10.1088/0004-637X/719/1/515}{\JournalTitle{\apj},
  719, 515}

\bibitem[{{Matt}(1994)}]{1994MNRAS.267L..17M}
{Matt}, G. 1994,
  \href{http://dx.doi.org/10.1093/mnras/267.1.L17}{\JournalTitle{\mnras}, 267,
  L17}

\bibitem[{{Miller} {et~al.}(2006){Miller}, {Raymond}, {Fabian}, {Steeghs},
  {Homan}, {Reynolds}, {van der Klis}, \& {Wijnands}}]{2006Natur.441..953M}
{Miller}, J.~M., {Raymond}, J., {Fabian}, A., {et~al.} 2006,
  \href{http://dx.doi.org/10.1038/nature04912}{\JournalTitle{\nat}, 441, 953}

\bibitem[{{Nayakshin} {et~al.}(2000){Nayakshin}, {Kazanas}, \&
  {Kallman}}]{2000ApJ...537..833N}
{Nayakshin}, S., {Kazanas}, D., \& {Kallman}, T.~R. 2000,
  \href{http://dx.doi.org/10.1086/309054}{\JournalTitle{\apj}, 537, 833}

\bibitem[{{Ponti} {et~al.}(2012){Ponti}, {Fender}, {Begelman}, {Dunn},
  {Neilsen}, \& {Coriat}}]{2012MNRAS.422L..11P}
{Ponti}, G., {Fender}, R.~P., {Begelman}, M.~C., {et~al.} 2012,
  \href{http://dx.doi.org/10.1111/j.1745-3933.2012.01224.x}{\JournalTitle{\mnras},
  422, 11}

\bibitem[{{Proga} \& {Kallman}(2002)}]{2002ApJ...565..455P}
{Proga}, D., \& {Kallman}, T.~R. 2002,
  \href{http://dx.doi.org/10.1086/324534}{\JournalTitle{\apj}, 565, 455}

\bibitem[{{Proga} {et~al.}(2000){Proga}, {Stone}, \&
  {Kallman}}]{2000ApJ...543..686P}
{Proga}, D., {Stone}, J.~M., \& {Kallman}, T.~R. 2000,
  \href{http://dx.doi.org/10.1086/317154}{\JournalTitle{\apj}, 543, 686}

\bibitem[{{Raymond}(1993)}]{1993ApJ...412..267R}
{Raymond}, J.~C. 1993,
  \href{http://dx.doi.org/10.1086/172917}{\JournalTitle{\apj}, 412, 267}

\bibitem[{{Ritter} \& {Kolb}(2003)}]{2003A&A...404..301R}
{Ritter}, H., \& {Kolb}, U. 2003,
  \href{http://dx.doi.org/10.1051/0004-6361:20030330}{\JournalTitle{\aap}, 404,
  301}

\bibitem[{{Schulz} {et~al.}(2008){Schulz}, {Kallman}, {Galloway}, \&
  {Brandt}}]{2008ApJ...672.1091S}
{Schulz}, N.~S., {Kallman}, T.~E., {Galloway}, D.~K., \& {Brandt}, W.~N. 2008,
  \href{http://dx.doi.org/10.1086/523809}{\JournalTitle{\apj}, 672, 1091}

\bibitem[{{Shakura} \& {Sunyaev}(1973)}]{1973A&A....24..337S}
{Shakura}, N.~I., \& {Sunyaev}, R.~A. 1973, \JournalTitle{\aap}, 24, 337

\bibitem[{{Shields} {et~al.}(1986){Shields}, {McKee}, {Lin}, \&
  {Begelman}}]{1986ApJ...306...90S}
{Shields}, G.~A., {McKee}, C.~F., {Lin}, D.~N.~C., \& {Begelman}, M.~C. 1986,
  \href{http://dx.doi.org/10.1086/164322}{\JournalTitle{\apj}, 306, 90}

\bibitem[{{Shlosman} \& {Vitello}(1993)}]{1993ApJ...409..372S}
{Shlosman}, I., \& {Vitello}, P. 1993,
  \href{http://dx.doi.org/10.1086/172670}{\JournalTitle{\apj}, 409, 372}

\bibitem[{{Shlosman} {et~al.}(1985){Shlosman}, {Vitello}, \&
  {Shaviv}}]{1985ApJ...294...96S}
{Shlosman}, I., {Vitello}, P.~A., \& {Shaviv}, G. 1985,
  \href{http://dx.doi.org/10.1086/163278}{\JournalTitle{\apj}, 294, 96}

\bibitem[{{Stone} \& {Norman}(1992)}]{1992ApJS...80..753S}
{Stone}, J.~M., \& {Norman}, M.~L. 1992,
  \href{http://dx.doi.org/10.1086/191680}{\JournalTitle{\apjs}, 80, 753}

\bibitem[{{Warner}(2003)}]{2003cvs..book.....W}
{Warner}, B. 2003, {Cataclysmic Variable Stars}

\bibitem[{{Woods} {et~al.}(1996){Woods}, {Klein}, {Castor}, {McKee}, \&
  {Bell}}]{1996ApJ...461..767W}
{Woods}, D.~T., {Klein}, R.~I., {Castor}, J.~I., {McKee}, C.~F., \& {Bell},
  J.~B. 1996, \href{http://dx.doi.org/10.1086/177101}{\JournalTitle{\apj}, 461,
  767}

\end{thebibliography}

\end{document}